\documentclass[%
preprint,
showpacs,
amsmath,amssymb,
aps,
]{revtex4-1}

\usepackage[dvipdfmx]{graphicx}
\usepackage{dcolumn}
\usepackage{bm}
\usepackage{setspace}

\usepackage{enumerate}

\begin{document}

\preprint{}

\title{
Universal function of the non-equilibrium phase transition
of nonlinear P\'{o}lya urn}

\author{Kazuaki Nakayama}
\email{nakayama@math.shinshu-u.ac.jp}
\affiliation{
Department of Mathematics,
Faculty of Science, Shinshu University, \\
Asahi 3-1-1, Matsumoto, Nagano 390-8621, Japan
}%

\author{Shintaro Mori}
\email{shintaro.mori@hirosaki-u.ac.jp}
\affiliation{
Department of Mathematics and Physics,
Faculty of Science and Technology, 
Hirosaki University, \\ 
Bunkyo-cho 3, Hirosaki, Aomori 036-8561, Japan
}

\date{\today}

\begin{abstract}
We study the phase transition and the critical properties  
of a nonlinear P\'{o}lya urn, which is a simple binary stochastic process
$X(t)\in \{0,1\},t=1,\cdots$ with a feedback mechanism. 
Let $f$ be a continuous function from the unit interval to itself, and
$z(t)$ be the proportion of the first $t$ variables $X(1),\cdots,X(t)$
that take the value 1. $X(t+1)$ takes the value 1 with probability $f(z(t))$.
When the number of stable fixed points of $f(z)$ changes,
the system undergoes a non-equilibrium phase transition and 
the order parameter is the limit
value of the autocorrelation function.
When the system is $Z_{2}$ symmetric, that is, $f(z)=1-f(1-z)$,
a continuous phase  transition occurs, and
the autocorrelation 
function behaves asymptotically as $\ln(t+1)^{-1/2}g(\ln(t+1)/\xi)$,
with a suitable definition of the correlation length $\xi$ and
the universal function $g(x)$.
We derive $g(x)$ analytically
using stochastic differential equations and
the expansion about the strength of stochastic noise.
$g(x)$ determines the asymptotic behavior of the autocorrelation
function near the critical point and the
universality class of
the phase transition.
\end{abstract}

\maketitle


\section{\label{sec:intro}Introduction}
Before the birth of econophysics \cite{Mantegna:2008},
Brian Arthur studied increasing returns or positive feedback
in economies and
demonstrated that they can magnify small perturbations in the market
\cite{Arthur:1989,Arthur:1990}.
 Let us assume that two selectively neutral
 technologies simultaneously enter the market.
 Owing to network externality, the utility of a
 product becomes an increasing function of its market share. 
 An initial small imbalance in the market share
 can eventually induce catastrophic imbalance.
 A similar mechanism also applies when people make decisions in an uncertain
 environment \cite{Bikchandani:1992}. In this case, it is rational
 to adopt the majority choice among previously chosen options, as
 this reflects the wisdom of the crowd  \cite{Surowiecki:2004}.
 The tendency to adopt the majority choice, overriding
 one's own private
 signal, is called information cascade or
 rational herding \cite{Bikchandani:1992}.
 Positive feedback from previous 
 choices affects later choices, and
 an initial small imbalance in the choices can
 have similar effects to those of increasing returns in the market.

 To describe the final catastrophic imbalance caused by the increasing returns
  and the information cascade, a non-linear P\'{o}lya urn model was
 adopted \cite{Hill:1980}.
 In the original P\'{o}lya urn model, an urn consists of $t$ balls,
 where the proportion of red balls is $z(t)\in (0,1)$, 
 and the rest of the balls are blue \cite{Polya:1931}.
 The probability of a new red ball being 
 added to the urn  
 is $z(t)$, whereas the corresponding probability for a blue ball is $1-z(t)$; 
 the proportion of red balls then becomes $z(t+1)$. 
 This procedure is iterative, and $z(t)$
 follows the beta distribution in the limit as $t\to\infty$.
 In nonlinear generalizations of this model, 
 a continuous function $f:[0,1]\to [0,1]$ determines the 
 probability $f(z(t)) $ of a red ball being added at stage $t+1$. 
 This nonlinear version is referred to as a nonlinear 
 P\'{o}lya process \cite{Hill:1980,Pemantle:1991}.
 In contrast to the original linear model, the nonlinear model 
 can have isolated stable states.
 The fixed point $z_{*}$ of $f(z)$, where $f(z_{*})=z_{*}$, 
 is (un)stable if $f'(z_{*})$ is smaller (greater) than 1 \cite{Hill:1980}.
 $z_{*}$ is referred to as downcrossing (upcrossing), 
 as the graph $y=f(z)$ crosses the curve $y=z$ in the  
 downward (upward) direction if $f'(z_{*})$ is smaller (larger)
 than 1. When  $f(z)$ touches the diagonal in the $(z,q)$ plane 
 at $z_{t}$, $z_{t}$ is referred to as the touchpoint.
 The stability of $z_{t}$ depends on the difference between the 
 slope of $f(z)$ and the diagonal $z$\cite{Pemantle:1991}. 

 The market share $z$ of a product
 determines the
 probability $f(z)$ that a new customer adopts it. 
 An S-shaped $f(z)$ function with two stable fixed points 
 suggests the random monopoly 
 formed when a technology 
 dominates over the other depends on initial chance fluctuations.
 This is a type of butterfly effect or high sensitivity
 to initial conditions in chaos theory; however, the final states
 are restricted to stable fixed states.
 An information cascade experiment
 provides a concrete setup for the physical realization of the
 formation of the final imbalance \cite{Anderson:1997}.
 The change in the number of stable states 
 can be observed by controlling the uncertainty
 of the subjects \cite{Mori:2012,Hino:2016}.

 In statistical physics,
 two aspects of non-linear P\'{o}lya urns have been studied.
 The super-normal transition in the
 convergence of $z(t)$ to a unique stable fixed point
 $z_{*}$ was studied in the context of long-range correlations in time series
 of financial data and DNA sequences 
 \cite{Hod:2004}.
 Another problem is to understand
 the change in the number of
 stable states as a non-equilibrium phase
 transition \cite{Mori:2015,Mori:2015-2}. The order parameter
 is the limit value of the autocorrelation function.
 If this value is zero, the memory of the past
 disappears in future variables. If it is
 positive, the memory of past variables remains and
 affects future variables forever.
 
 The critical behavior
 of the autocorrelation function bears a strong resemblance to
 the order parameters of absorbing-state phase transitions
 \cite{Hinrichsen:2000}.
 If $f(z)$ is given as the superposition of
 a constant function and a step function as
 $(1-p)q+p\theta(z-1/2),q\neq 1/2,0<p<1$ (we term this the digital model),
 a continuous phase transition occurs at
 $p=p_{c}(q)=1-1/2q$\cite{Hisakado:2011}.
 For $p<p_{c}(q)$, there is only one stable
 state at $z_{*}=(1-p)q+p$, and the correlation function decays exponentially.
 At the critical point $p=p_{c}(q)$, where there is a stable fixed point
 at $z_{*}=(1-p)q+p$ and
 an unstable touchpoint at $z_{t}=(1-p)q$,
 the autocorrelation function exhibits power-law decay.
 For $p>p_{c}(q)$, there are two stable states $z_{\pm}$ at
 $(1-p)q+p,(1-p)q$, and  the order parameter
 becomes positive \cite{Mori:2015}.
 Furthermore, the asymptotic behavior of the autocorrelation function
 obeys a scaling law, namely, $b(q)t^{-1/2}g(t/\xi(q,p))$, with
 universal function $g(x)$ \cite{Mori:2015-2}.
 If $f(z)$ is a smooth function, the system undergoes a continuous phase
 transition when $f(z)=1-f(1-z)$.
 The critical and the off-critical behavior of the autocorrelation function
 are completely different from those of the aforementioned digital model.
 At the critical point, the autocorrelation
 function decays as $\sim \ln t^{-1/2}$.
 Below the critical point, the autocorrelation function
 exhibits power-law decay in $t$, and the power-law exponent
 is determined as $f'(z_{*})-1$.
 Based on the behaviors
 and the analogy with the digital model,
 we propose the scaling form
 $b(\ln t)^{-1/2}g(\ln t/\xi)$ for the autocorrelation function.
 Here, we define the correlation length
 $\xi$ using the exponent of the power-law decay as $\xi=1/(1-f'(z_{*}))$.

 In this study, we derive the universal function $g(x)$ 
 in the case where  $f(z)$ is a smooth function as above.
 We map the stochastic
 process to a stochastic differential
 equation (SDE) and expand its trajectory about the strength of the noise.
 We solve the initial value problem 
 and analytically derive the autocorrelation function.
 Then, we derive $g(x)$ and verify it numerically. 
 The paper is organized as follows.
 We define the model in Section \ref{sec:model}, and map
 it to an SDE and expand its trajectory about the strength of the
 noise in Section \ref{sec:analysis}.
 In Section \ref{sec:result}, we solve the corresponding initial value problem, estimate the asymptotic behavior of the autocorrelation function,
 and derive the universal function.
 In Section \ref{sec:result2}, we numerically estimate the autocorrelation and
 the universal function, and we verify that the
 derived universal function completely describes the numerical results.
 We summarize the results in Section \ref{sec:con}.

\section{\label{sec:model}Model}
 
We define a binary stochastic process 
$X(t)\in \{0,1\},t\in \{1,2,\cdots,T\}$, 
where the probability 
that $X(t)$ takes the value 1 is given by a function $f(z,h)$
of the proportion  $z(t-1)$ of the variables $X(1),\cdots,X(t-1)$
that are equal to 1. 
\begin{eqnarray}
f(z,h)&\equiv& \mbox{P}(X(t)=1|z(t-1)=z)=z-(z-1/2)^3+p(z-1/2)+h
\nonumber \\
z(t)&=&\frac{1}{t}\sum_{s=1}^{t}X(s)\,\,\,\, \mbox{for} \,\,\,\, t>0 , \,\, \mbox{and}\,\, 
z(0)=\frac{1}{2} 
\label{eq:model}.
\end{eqnarray}
The choice of $f(z,h)$ is arbitrary, and we adopt the above form.
In the numerical studies, we adopt another
form for $f(z,h)$ to verify that the results are independent
of the choice of $f(z,h)$.
With this choice for $f(z,h)$, we can
estimate the fixed point of $f(z,h)$ by solving $f(z)=z$ explicitly.
Here, we consider the parameter space $(p,h)$, where
$0\le f(z,h)\le 1$ for $\forall 0\le z\le 1$.

\begin{figure}[htbp]
\begin{center}
\includegraphics[width=7cm]{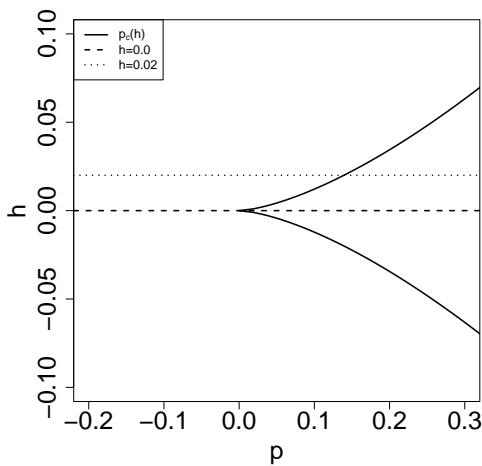}
\end{center}
\caption{
Phase diagram in $p-h$ plane.  
The phase boundary is given by \eqref{eq:pc_h}.
For $p<p_{c}(h)$, there is a stable fixed point at $z_{*}$.
For $p>p_{c}(h)$, there are two stable fixed points at $z_{\pm}$.
For $h=0,p=p_{c}(0)=0$, there is a touchpoint at $z=0$.
For $h\neq 0$ and $p=p_{c}(h)$, there is a stable fixed point at
$z_{*}$, and a touchpoint at $z_{t}$.
}
\label{fig:p_h}
\end{figure}

The number of stable fixed points depends on $(p,h)$, and there is a threshold value $p=p_{c}(h)$, which is a function of $h$ 
(Fig. \ref{fig:p_h}):
\begin{equation}
p_{c}(h)=3(h/2)^{2/3} \label{eq:pc_h}
\end{equation}  
For $p<p_{c}(h)$, there is only one fixed point 
at $z=z_{*}$. 
As $p$ increases and $h$ is fixed, $f(z,h)$ becomes tangential
to the diagonal at $z_{t}$ for $p=p_{c}(h)$. 
For $h\neq 0$, $z_{t}\neq z_{*}$, and both $z_{t}$ and $z_{*}$ 
are stable. 
For $h=0$, $z_{t}$ and $z_{*}$ are equal and stable.
In both cases, the slope of the curve at $z_{t}$ is equal to 1.
For $p>p_{c}(h)$, there are three fixed points, and we denote them 
as $z_{-}<z_{u}<z_{+}$; $z_{-}$ and $z_{+}$ are stable, whereas $z_{u}$ is unstable.
We denote the slope of $f(z,h)$ at $z_{*}$ and $z_{\pm}$ by
$l_{*}$ and $l_{\pm}$, respectively. As $z_{*}$ and $z_{\pm}$ are stable and downcrossing,
$l_{*},l_{\pm}<1$.

The autocorrelation function $C(t)$ (correlation function
for brevity) is defined as the  difference 
of the two conditional probabilities on $X(1)$:
\begin{equation}    
C(t)=\mbox{P}(X(t+1)=1|X(1)=1)-\mbox{P}(X(t+1)=1|X(1)=0) 
\label{eq:Ct}.
\end{equation}
$C(t)$ can be defined as $\mbox{Cov}(X(1),X(t+1))/\mbox{V}(X(1))$ using the variance of $X(1)$ and the covariance of $X(1)$ and $X(T+1)$.
The asymptotic behavior of $C(t)$ depends on $(p,h)$.
In the following, we analytically derive this behavior by mapping the
model to an SDE.

\section{\label{sec:analysis}Model Analysis}

\subsection{Stochastic Differential Equation}

The random variable $z(t)$ defined in
\eqref{eq:model}
satisfies the following recursion relation:
\begin{equation}
  \label{eq:z.recursion}
  z(t+1) = z(t) + \frac{X(t+1) - z(t)}{t+1}.
\end{equation}
The conditional expectation and the conditional variance of $z(t)$
are estimated as follows:
\begin{subequations}
\begin{align}
  E(z(t+1)|z(t)=z) &= z + \frac{f(z,h) - z}{t+1},
  \label{eq:cond.drift}
  \\
  V(z(t+1)|z(t)=z) &= \frac{f(z,h)\{1-f(z,h)\}}{(t+1)^{2}}.
  \label{eq:cond.var}
\end{align}
\end{subequations}
The second term on the right-hand side of 
\eqref{eq:cond.drift} is regarded as a drift term,
whereas the right-hand side of \eqref{eq:cond.var} is interpreted as a
diffusion coefficient.
Thus, we introduce the continuous-time model described by
the following Ito-type SDE:
\begin{equation}
  \label{eq:SDE0}
  dz = \frac{f(z,h)-z}{t+1}\,dt + \frac{\sqrt{f(z,h)\{1-f(z,h)\}}}{t+1}\,dW_{t},
\end{equation}
where $W_{t}$ is the Wiener process \cite{Gardiner:2009}.
In the following, we study the following simplified SDE model:
\begin{equation}
  \label{eq:SDE1}
  dz = \frac{f(z,h)-z}{t+1}\,dt + \frac{D}{t+1}\,dW_{t},
\end{equation}
where $D$ is a small positive constant.
It seems that there is no significant difference between 
\eqref{eq:SDE0} and \eqref{eq:SDE1}
as far as the long-time behavior of the system is
concerned\footnote{This is true at least up to $O(D^{2})$ because
the classical solutions $y_{0}^{\pm}(t)$ respectively converge to $a_{\pm}$.}.

\subsection{Small Noise Approximation}

Let $y=z-1/2$ be a new random variable.
We define
$g(y,h)=\left[f(z,h) - z\right]_{z=y+1/2} = -y^{3} + py + h$.
Then, Eq.~\eqref{eq:SDE1} becomes
\begin{equation}
  \label{eq:SDE2}
  dy = \frac{g(y,h)}{t+1}\,dt + \frac{D}{t+1}\,dW_{t}.
\end{equation}
We expand $y(t)$ in powers of $D$:
\begin{equation}
  \label{eq:yD}
  y(t) = y_{0}(t) + Dy_{1}(t) + D^{2}y_{2}(t) + \cdots.
\end{equation}
Then, Eq.~\eqref{eq:SDE2} is solved recursively.
As we are interested in the correlation function $C(t)$,
we adopt the initial condition $y(0)=\pm1/2$, that is,
\begin{equation}
  \label{eq:init}
  y_{n}(0) = \pm\frac{1}{2}\delta_{n0}\quad(n=0,1,2,\ldots).
\end{equation}
For simplicity, $y^{\pm}(t)$ and $y^{\pm}_{n}(t)$ denote the
solutions corresponding to each initial condition.

We summarize the solution of the initial value problem
corresponding to \eqref{eq:SDE2}.
The details of the calculation are given in Appendix~\ref{sec:SNA}.
We first note that the classical solution $y_{0}(t)$ of \eqref{eq:SDE2}
is independent of $D$.
It is easily obtained as an implicit function:
\begin{equation}
  \label{eq:y0}
  \ln(t+1) = \int_{\pm1/2}^{y_{0}^{\pm}}\!\frac{dy}{g(y,h)}.
\end{equation}
Furthermore, the expectation value and the variance of $y(t)$
are expressed, up to the second order of $D$, as
\begin{subequations}
  \begin{align}
    \label{eq:Ey}
    E(y^{\pm}(t)) &= y_{0}^{\pm}(t) + D^{2}E(y_{2}^{\pm}(t)),
    \\
    \label{eq:Vy}
    V(y^{\pm}(t)) &= D^{2}E(y_{1}^{\pm}(t)^{2}).
  \end{align}
\end{subequations}
Finally, these expectation values are explicitly obtained as follows:
\begin{subequations}
  \begin{align}
    \label{eq:y1}
    E(y_{1}^{\pm}(t)^{2})
    &= e^{2G^{\pm}(t)}\int_{0}^{t}\!\frac{e^{-2G^{\pm}(t)}}{(t+1)^{2}}\,dt
    \\
    \label{eq:y2}
    E(y_{2}^{\pm}(t))
    &= -3e^{G^{\pm}(t)}\int_{0}^{t}\!\frac{e^{-G^{\pm}(t)}y_{0}^{\pm}(t)E(y_{1}^{\pm}(t)^{2})}{t+1}\,dt,
    \\
    \intertext{where}
    \label{eq:G}
    G^{\pm}(t) &= p\ln(t+1) - 3\int_{0}^{t}\!\frac{x_{0}^{\pm}(t)^{2}}{t+1}\,dt.
  \end{align}
\end{subequations}

\section{\label{sec:result}Correlation function and universal function}

\subsection{Correlation Function and Order Parameter}

The correlation function $C(t)$ (defined in \eqref{eq:Ct})
is expressed as \cite{Mori:2015-2}
\begin{equation}
  \label{eq:corr3}
  C(t) = E(y^{+}(t)+g(y^{+}(t),h)) - E(y^{-}(t)+g(y^{-}(t),h)).
\end{equation}
Let $a_{-}\le a_{+}$ be the stable fixed points of $g(y,h)$.
The order parameter $c=c(p,h)$ is the difference $a_{+} - a_{-}$ of
the stable fixed points because
$y^{\pm}(t) \rightarrow a_{\pm}$
as
$t\rightarrow\infty$
up to $O(D^{2})$.

Our result is that
\begin{equation}
  \label{eq:p_vs_c}
  c(p,h) =
  \begin{cases}
    2\sqrt{p}\cos\dfrac{\pi-2\mu(p,h)}{6}, & p\ge p_{c}(h)=3(h/2)^{2/3},
    \\
    0, & p< p_{c}(h)=3(h/2)^{2/3},
  \end{cases}
\end{equation}
where
\begin{equation}
  \label{eq:mu1}
  \mu(p,h)
  =
  \begin{cases}
    \arccos\left\{\dfrac{|h|}{2}\left(\dfrac{3}{p}\right)^{3/2}\right\}, & h\neq0,
    \\
    \dfrac{\pi}{2}, & h=0.
  \end{cases}
\end{equation}
Figure~\ref{fig:p_h} shows the phase diagram in the $p-h$ plane.
The ordered phase with $c(p,h)>0$ is the region where
$p \ge  p_{c}(h)=2(p/3)^{3/2}$.

The correlation function $C(t)$ is
\begin{subequations}
  \label{eq:corr2}
\begin{align}
  C(t)
  &= C_{0}(t) + D^{2}C_{2}(t),
  \\
  C_{0}(t)
  &= y_{0}^{+}(t) - y_{0}^{-}(t) + g(y_{0}^{+}(t),h)
  - g(y_{0}^{-}(t),h),
  \\
  C_{2}(t)
  &= E(y_{2}^{+}(t)) - E(y_{2}^{-}(t))
  + \{g_{y}(y_{0}^{+}(t),h)E(y_{2}^{+}(t))
  - g_{y}(y_{0}^{-}(t),h)E(y_{2}^{-}(t))\}
  \notag
  \\
  &\qquad
  + \frac{1}{2}\{g_{yy}(y_{0}^{+}(t),h)E(y_{1}^{+}(t)^{2})
  - g_{yy}(y_{0}^{-}(t),h)E(y_{1}^{-}(t)^{2})\},
\end{align}
\end{subequations}
where $C_{0}(t)$ and $C_{2}(t)$ represent the $O(D^{0})$ and
$O(D^{2})$ parts of the function $C(t)$, respectively.

\subsection{Correlation Function in the Non-symmetric Domain}

We summarize the asymptotic behavior of
the correlation function when $h>0$.

\paragraph{Ordered Phase}

In the off-critical region $p>3(h/2)^{2/3}$,
the correlation function exhibits power-law decay:
\begin{equation}
  \label{eq:corr.ph.1}
  C(t) \sim c(p,h) + c^{\prime}(t+1)^{-\zeta},
\end{equation}
where $c^{\prime}$ is a constant, and
\begin{equation}
  \label{eq:zeta1}
  \zeta = p + 2p\sin\frac{4\mu(p,h)-\pi}{6}.
\end{equation}
The function $\mu(p,h)$ is defined in
\eqref{eq:mu1}.

On the critical line $p=3(h/2)^{2/3}$, the function exhibits logarithmic behavior:
\begin{equation}
  \label{eq:corr.ph.2}
  C(t) \sim \sqrt{3p} + \frac{1}{\sqrt{3p}\ln(t+1) + c^{\prime}}.
\end{equation}

\paragraph{Disordered Phase}

When $p<3(h/2)^{2/3}$, the correlation function behaves as follows:
\begin{equation}
  \label{eq:corr.ph.3}
  C(t) \sim c^{\prime}\cdot
  \begin{cases}
    (t+1)^{-\zeta}, & \zeta\neq1,
    \\
    (t+1)^{-2}\ln(t+1), & \zeta=1,
  \end{cases}
\end{equation}
where
\begin{equation}
  \label{eq:zeta2}
  \zeta =
  \begin{cases}
    p\left(3\cosh^{2}\frac{\mu}{3} + \sinh^{2}\frac{\mu}{3}\right), & p>0,
    \\
    |p|\left(3\sinh^{2}\frac{\mu}{3} + \cosh^{2}\frac{\mu}{3}\right), & p<0,
    \\
    3h^{2/3}, & p=0,
  \end{cases}
\end{equation}
and the parameter $\mu=\mu(p,h)$ is given by
\begin{equation}
  \label{eq:mu2}
  \mu(p,h) =
  \begin{cases}
    \cosh^{-1}\left\{\dfrac{h}{2}\left(\dfrac{3}{p}\right)^{3/2}\right\}, & p>0,
    \\
    \sinh^{-1}\left\{\dfrac{h}{2}\left(\dfrac{3}{|p|}\right)^{3/2}\right\}, & p<0.
  \end{cases}
\end{equation}

\subsection{Correlation Function in the  Symmetric  Domain}

When $h=0$, exact solutions can be obtained.
The off-critical correlation function is estimated as
\begin{subequations}
  \label{eq:corr.p}
  \begin{align}
    \label{eq:corr.p1}
    C(t;p) &= 2\sqrt{\frac{p}{1-(1-4p)(t+1)^{-2p}}} + O(D^{2}),
    \quad
    p\neq0.
    \\
    \intertext{whereas the on-critical correlation function is given by}
    \label{eq:corr.p2}
    C(t;0) &= \frac{1}{\sqrt{(1/2)\ln(t+1)+1\}}} + O(D^{2}).
  \end{align}
\end{subequations}
The function $C(t;p)$ is continuous, that is,
$\displaystyle{\lim_{p\rightarrow\pm0}C(t;p)}$
coincides with $C(t;0)$.

\subsection{Universal Function}
In the ordered phase, the asymptotic behavior of the correlation function
\eqref{eq:corr.p1}
is $C(t)-c(p,0)\sim(t+1)^{-2p}=e^{-2p\ln (t+1)}$.
The correlation length $\xi$ on the $\ln(t+1)$ scale is $1/2p$
and diverges at $p=0$.
Therefore, keeping $x=2p\ln(t+1)$ constant and taking the limit
as $\ln(t+1),\xi=1/2p \to \infty$,
we have the universal function
\begin{subequations}
  \label{eq:x_vs_UF}
\begin{equation}
  \label{eq:x_vs_UF.1}
  g(x) = \lim_{p\rightarrow+0}\frac{C(t;p)}{C(t;0)}
  = \sqrt{\frac{x}{1-e^{-x}}}.
\end{equation}
By contrast, in the disordered phase, 
the asymptotic behavior of the correlation function
\eqref{eq:corr.p2} is
$C(t)\sim(t+1)^{-|p|}=e^{-|p|\ln (t+1)}$. Therefore,
$x=|p|\ln(t+1)$ should be constant to estimate the universal function.
Thus, 
\begin{equation}
  \label{eq:x_vs_UF.2}
  g(x) = \lim_{p\rightarrow-0}\frac{C(t;p)}{C(t;0)}
  = \sqrt{\frac{2x}{e^{2x}-1}}.
\end{equation}
\end{subequations}

In the ordered phase ($p>0$), $g(x)\propto x^{1/2}$ for large $x$.
As $C(t)=C(t;0)g(2p\ln (t+1))$ and $C(t;0)\propto \ln(t+1)^{-1/2}$ for
$\ln(t+1)>>1$, we have $C(t)\propto p^{1/2}$.
The critical exponent $\beta$ for the
order parameter $c\propto p^{\beta}$ is $1/2$, which
is suggested based on the scaling hypothesis
for $C(t)$ \cite{Mori:2015-2}.

We also consider the universal function in the $h$ direction.
Let $C(t;p,h)$ be the correlation function.
By \eqref{eq:corr.ph.3}, we have $C(t;0,h)\sim c^{\prime}(t+1)^{-3h^{2/3}}$
when $h>0$ is small.
Thus, we take the limit as $\ln(t+1), \xi=1/3h^{2/3} \rightarrow \infty$
and keep the quantity $x=3h^{2/3}\ln(t+1)$ constant.
Then,
\begin{subequations}
    \label{eq:xh_vs_UF}
\begin{equation}
  \label{eq:xh_vs_UF.1}
  g_{h}(x) = \lim_{h\rightarrow+0}\frac{C(t;0,h)}{C(t;0,0)}
  = \frac{\sqrt{x}}{2\sqrt{2}}\left\{
    \tan\varphi^{-1}\left(e^{-x}\varphi(\pi/2)\right)
    - \tan\varphi^{-1}\left(e^{-x}\varphi(-\pi/2)\right)
  \right\},
\end{equation}
where the monotonically increasing function $\varphi(\theta)$ is defined as follows:
\begin{equation}
  \label{eq:xh_vs_UF.2}
  \varphi(\theta) = e^{-\sqrt{3}\theta}\sin\left(\theta-\frac{\pi}{3}\right),\quad
  |\theta|\le\frac{\pi}{2}.
\end{equation}
\end{subequations}

\section{\label{sec:result2}Numerical Study of Phase transition}

\begin{figure}[htbp]
\begin{center}
\includegraphics[width=10cm]{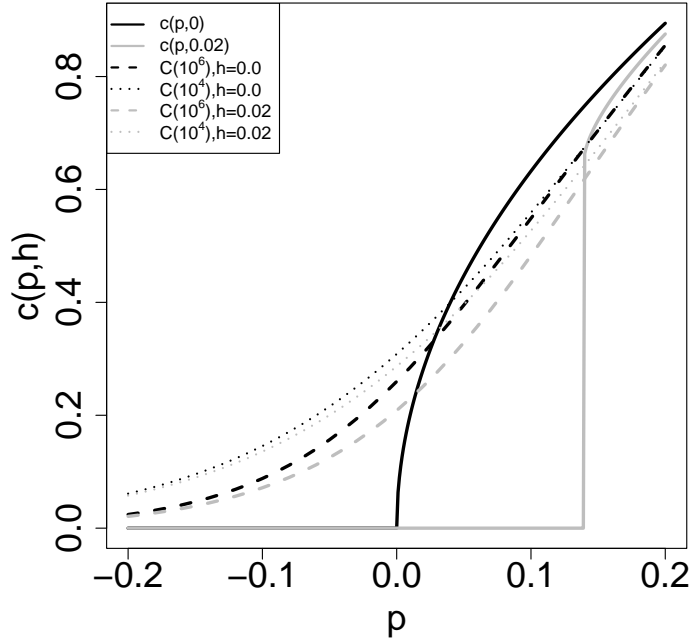}
\end{center}
\caption{
  Plot of $C(T=10^6|p,h)$ vs. $p$.
  We set $h=0.0(\Box),0.02(\circ)$
  and $p\in [-0.2,0.2]$.
The result of Eq.~\eqref{eq:p_vs_c} is also plotted with
solid and dotted curves.
}
\label{fig:p_vs_c}
\end{figure}

We perform numerical integration of the master equation  
and estimate $C(t)$.  
We denote the probability function for 
$\sum_{s=1}^{t}X(s)$ with initial condition $X(1)=x_1$ as
$P(t,n|x_{1})\equiv \mbox{P}(\sum_{s=1}^{t}X(s)=n|X(1)=x_{1})$.
We have that $P(1,n|x_1)=\delta_{n,x_1}$ holds.
The master equation for $P(t,n|x_{1})$ is
\begin{equation}
P(t+1,n|x_{1})=f((n-1)/t,h)\cdot P(t,n-1|x_{1})
+(1-f(n/t,h))\cdot P(t,n|x_{1}).
\end{equation}
We impose the boundary conditions 
$P(t,n,x_{1})=0$ for $n<0$ or $n>t$.
Using $P(t,n|x_1)$, we  estimate $C(t)$ for $t\le T=10^6$ as
\[
C(t|p,h)=\sum_{n=0}^{t}f(n/t,h)(P(t,n|1)-P(t,n|0)).
\]
We set $h\in \{0.0,0.02\}$ and $p\in [-0.2,0.2]$.
The plot of $C(T|p,h)$ vs. $p$ as well as the plot of $c(p,h)$ from Eq.~\eqref{eq:p_vs_c} are shown in Fig. \ref{fig:p_vs_c}.

It is seen that $C(T|p,h)$ for $T=10^6$ is rather different
from $c(p,h)=\lim_{T\to\infty}C(T|p,h)$ near $p=p_{c}(h)$.
One cannot observe the transition between the phase with
$c=0$ for $p<p_{c}(h)$
and that with $c>0$ for $p>p_{c}(h)$. This discrepancy can be accounted for by the
limited system size $T$ and the strong correlation
near $p=p_{c}(h)$. To see the thermodynamic limit
$T\to \infty$ and the phase transition, it is necessary to study
the scaling function $g(x)$.

We numerically estimate $g(x)$ defined
in Eq.~\eqref{eq:x_vs_UF}. We estimate $C(T|p,0)/C(t|0,0)$
for $\xi(p)>>1$ as a function of $x=T/\xi(p)$.
We set $p\in [-0.01,0.01]$,
$[-0.001,0.001]$,
$[-0.0001,0.0001]$.
We also estimate $x(p)\equiv T/\xi(p)$ using Eq.~\eqref{eq:p_vs_xi}.
\begin{equation}
\xi(p)=
\left\{
\begin{array}{cc}
1/|p| &  p<0 \\
1/2p    &  p>0 
\end{array}  
\right.  \label{eq:p_vs_xi}
\end{equation}

In addition to the model in Eq.~\eqref{eq:model},
we also study another model with the following function:
\begin{equation}
f(z)=\frac{1}{2}\left\{\tanh(2J(z-1/2)+h)+1)  \right\}
\label{eq:model2}  
\end{equation}
A continuous phase transition occurs at $h=0,J=1$.
For $h=0$ and $J<1$, $z_{*}=1/2$ and $l_{*}=f'(z_{*})=J$.
For $h=0$ and $J>1$, there are stable states $z_{\pm}$ and
$l_{\pm}=J/\cosh^2(2J(z_{\pm}-1/2))$. The correlation length $\xi(J,0)$
is defined as
\begin{equation}
\xi(J)=
\left\{
\begin{array}{cc}
1/(1-J) &  J<1 \\
1/(1-J/\cosh^2(2J(z_{\pm}-1/2))   &  J>1 
\end{array}  
\right.\label{eq:p_vs_xi2}
\end{equation}
We set $J\in [0.99,1.01]$,$[0.999,1.001]$,$[0.9999,1.0001]$
and estimate $C(T|J,0)/C(T|1,0)$ for $T=10^6$.
We also estimate $x(J)\equiv T/\xi(J)$ using Eq.~\eqref{eq:p_vs_xi2}.

\begin{figure}[htbp]
\begin{center}
\begin{tabular}{cc}    
\includegraphics[width=8cm]{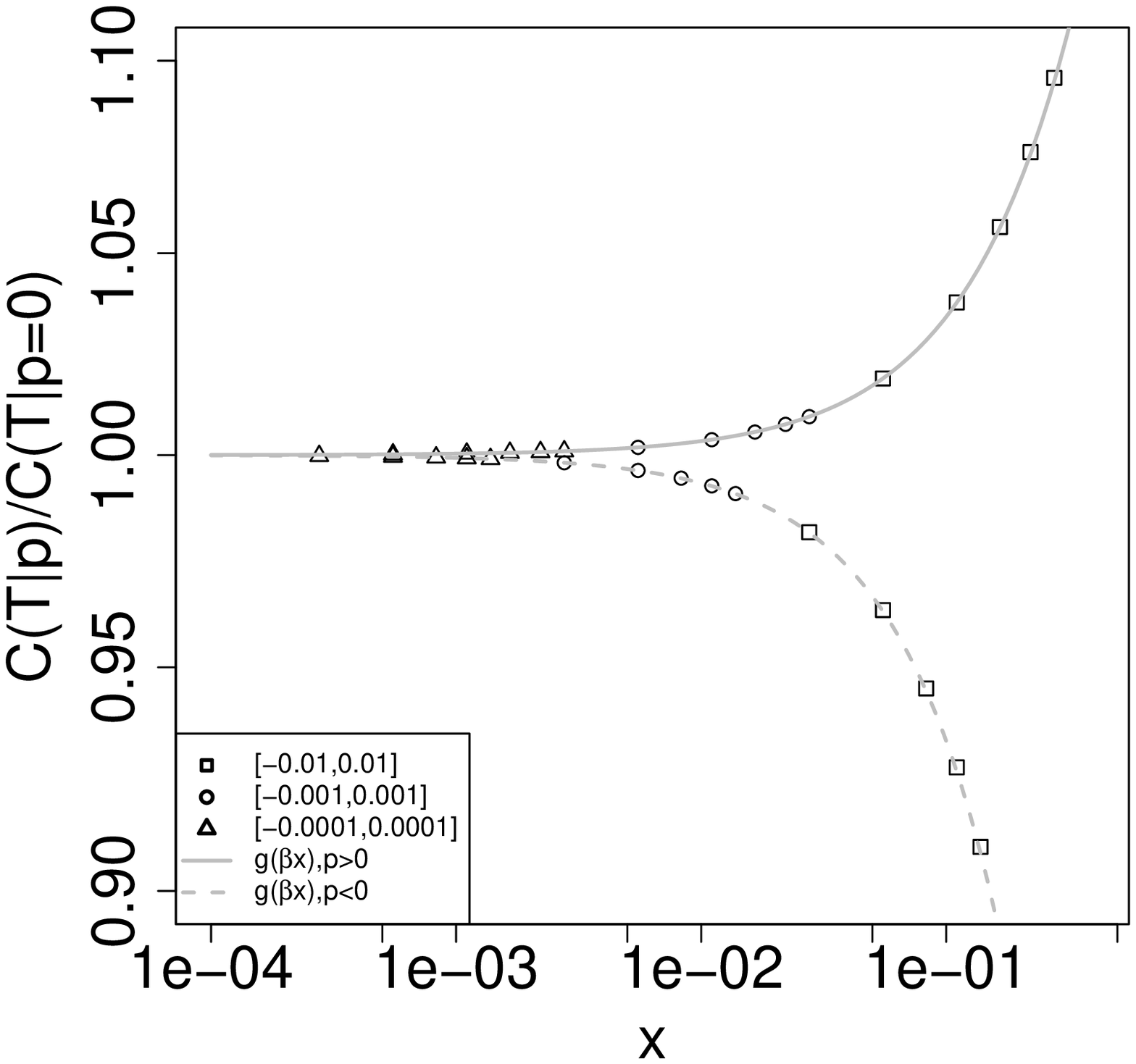}
&   
\includegraphics[width=8cm]{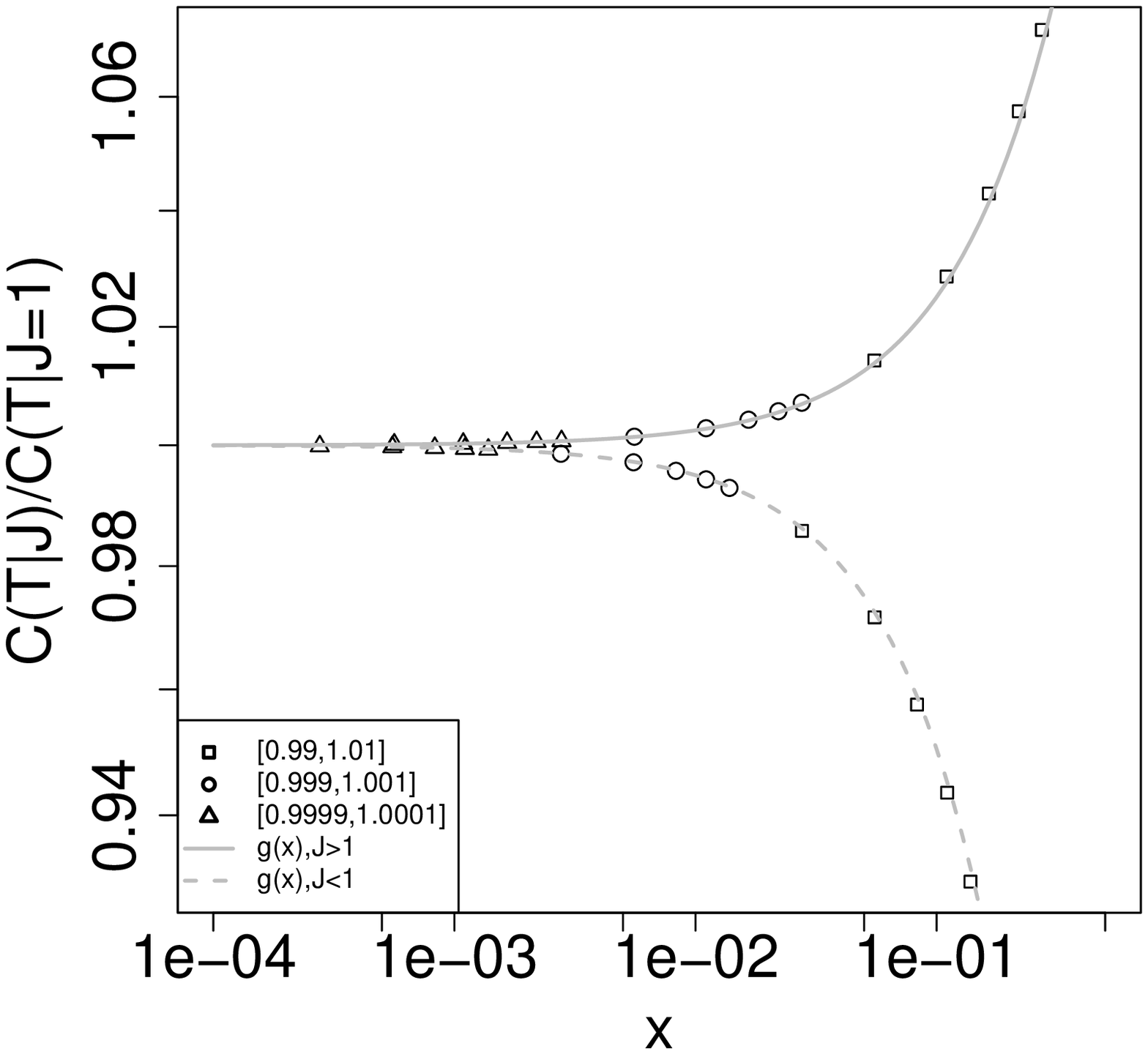}
\end{tabular}
\end{center}
\caption{
(a) Plot of $C(T|p,0)/C(T|0,0)$ vs. $x=T/\xi(p)$.
We set 
$p\in [-0.01,0.01](\Box)$, $[-0.001,0.001](\circ)$,
$[-0.0001,0.0001](\triangle)$, $h=0$ and $T=10^6$;
$g(1.35x)$ from Eq.~\eqref{eq:x_vs_UF} is plotted with
gray solid $(p>0)$ and broken curves $(p<0)$.
(b) Plot of $C(T|J,0)/C(T|1,0)$ vs. $x=T/\xi(J)$.
We set
$J\in [-0.99,1.01](\Box)$, $[-0.999,1.001](\circ)$,
$[0.9999,1.0001](\triangle)$, $h=0$ and $T=10^6$;
$g(x)$ from Eq.~\eqref{eq:x_vs_UF} is plotted with
gray solid $(J>1)$ and broken curves $(J<1)$.
}
\label{fig:x_vs_UF}
\end{figure}

The plots of $C(T|p,0)/C(T|0,0)$ vs. $x(p)$
and $C(T|J,0)/C(T|1,0)$ vs. $x(J)$
are shown in Fig. \ref{fig:x_vs_UF}.
It is observed that the numerical results are well described by 
Eq.~\eqref{eq:x_vs_UF}. It can be clearly seen that
there are two phases: $c>0$ and $c=0$.
When $p<0$, or $J<1$, $g(x)$ decays to zero for large $x$.
As $C(T|p,0)\simeq C(T|0,0)g(x=\ln (T+1)/\xi(p,0)$,
we have $c=\lim_{T\to \infty}C(T|p,0)=0$.
When $p>0$, $g(x)$ grows as $x^{1/2}$ for large $x$.
Then, as $C(T|0,0)\simeq b\ln(T+1)^{-1/2}$, we have
$c=\lim_{T\to \infty}C(T|p,0)=b\ln(T+1)^{-1/2}g(\ln (T+1)/\xi(p))
\propto \xi(p)^{-1/2}>0
$.

\begin{figure}[htbp]
\begin{center}    
\includegraphics[width=8cm]{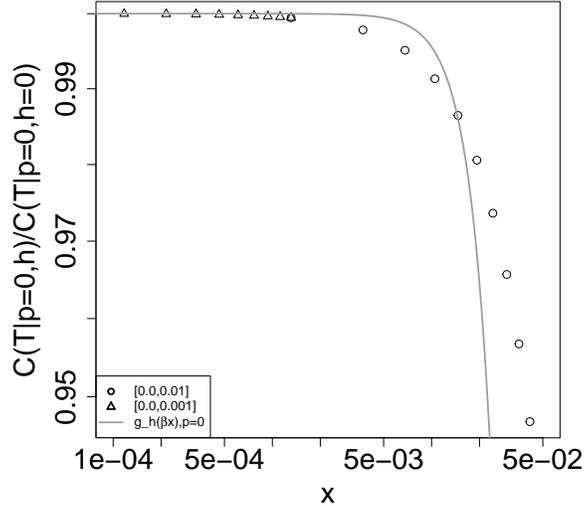}
\end{center}
\caption{
Plot of $C(T|0,h)/C(T|0,0)$ vs. $x=T/\xi(h)$.
We set 
$h\in [0.0,0.01](\circ)$, $[0,0.001](\triangle)$, $p=0$ and $T=10^6$;
$g_{h}(10^2x)$ from Eq.~\eqref{eq:xh_vs_UF} is plotted with
gray solid line.
}
\label{fig:xh_vs_UF}
\end{figure}

We also check the universal function in the $h$ direction
in Fig. \ref{fig:xh_vs_UF}.
We adopt the same procedures as in the $p$ direction, 
and we estimate $C(T|0,h)/C(T|0,0)$ for $T=10^6$ and
$h\in [0.0,0.01]$,$[0.00,0.001]$. 
It can be seen that $g_{h}(x)$ in Eq.~\eqref{eq:xh_vs_UF} does not describe
$C(T|0,h)/C(T|0,0)$. The symmetry breaking field $h$ disrupts the
universal asymptotic behavior of $C(T|h,0)$.

\section{\label{sec:con}Summary}

We studied the 
phase transition of a nonlinear P\'{o}lya
urn using an SDE. We solved the initial value
problem  and analytically estimated the correlation function $C(T)$. By taking the scaling limit as $T\to \infty$
and $\xi\to\infty$, with $x=T/\xi$ fixed,
we derived the universal function of the phase transition,
which governs the asymptotic behavior of $C(T)$ near the critical point.
We also numerically verified the results, 
where the phase transition
of the system could be clearly observed.

Our study ascertained that scaling analysis is a powerful method
to understand the non-equilibrium phase transition of
non-linear P\'{o}lya urns.
In addition, SDEs and the expansion about the strength of noise are also 
useful to understand
the critical and universal behavior in non-equilibrium phase transition.
A non-linear P\'{o}lya urn is one of the
simplest systems in which time evolution is greatly affected by feedback.
In nature, there are many stochastic processes where some feedback mechanism
plays crucial role, as Hawks process etc\cite{Hawks:1971,Kanazawa:2020}.
We believe the scaling analysis continues to be promising approaches
to understand the universal behavior of such systems.

\begin{acknowledgments}
The authors thanks Yugo Kagaya for useful discussions.  
This work was supported by JPSJ KAKENHI[Grant No. 17K00347].  
\end{acknowledgments}

\providecommand{\noopsort}[1]{}\providecommand{\singleletter}[1]{#1}%

\appendix

\section{Small Noise Approximation}
\label{sec:SNA}

By substituting \eqref{eq:yD} into \eqref{eq:SDE2},
a sequence of SDEs is obtained:
\begin{subequations}
  \begin{align}
    \label{eq:SDEy0}
    dy_{0} &= \frac{g(y_{0},h)}{t+1}\,dt,
    \\
    \label{eq:SDEy1}
    dy_{1} &= \frac{g_{y}(y_{0},h)y_{1}}{t+1}\,dt
    + \frac{1}{t+1}\,dW_{t},
    \\
    \label{eq:SDEy2}
    dy_{2} &= \frac{g_{y}(y_{0},h)y_{2} + \frac{1}{2}g_{yy}(y_{0},h)y_{1}^{2}}{t+1}\,dt.
  \end{align}
\end{subequations}
Equation \eqref{eq:SDEy0} immediately yields Eq.~\eqref{eq:y0}.
We now define $u_{k}^{\pm}(t)=e^{-G^{\pm}(t)}y_{k}^{\pm}(t)$,
$k=1,2$, where $G^{\pm}(t)$ was defined in Eq.~\eqref{eq:G}.
Then, Eqs.~\eqref{eq:SDEy1} and \eqref{eq:SDEy2} become
\begin{subequations}
  \begin{align}
    du_{1}^{\pm} &= \frac{e^{-G^{\pm}(t)}}{t+1}\,dW_{t},
    \\
    du_{2}^{\pm} &= -3\frac{e^{-G^{\pm}(t)}y_{0}^{\pm}(t)y_{1}^{\pm}(t)^{2}}{t+1}\,dt.
  \end{align}
\end{subequations}
Thus, we have the closed-form solutions
\begin{subequations}
\begin{align}
  y_{1}^{\pm}(t) &= e^{G^{\pm}(t)}\int_{0}^{t}\!\frac{e^{-G^{\pm}(t)}}{t+1}\,dW_{t},
  \\
  y_{2}^{\pm}(t) &= -3e^{G^{\pm}(t)}\int_{0}^{t}\!\frac{e^{-G^{\pm}(t)}y_{0}^{\pm}(t)y_{1}^{\pm}(t)^{2}}{t+1}\,dt.
\end{align}
\end{subequations}
The expectation values are estimated as
\begin{subequations}
  \begin{align}
    E(y_{1}^{\pm}(t)) &= 0,
    \\
    E(y_{2}^{\pm}(t)) &= -3e^{G^{\pm}(t)}\int_{0}^{t}\!\frac{e^{-G^{\pm}(t)}y_{0}^{\pm}(t)E(y_{1}^{\pm}(t)^{2})}{t+1}\,dt,
    \\
    E(y_{1}^{\pm}(t)^{2}) &= e^{2G^{\pm}(t)}\int_{0}^{t}\!\frac{e^{-2G^{\pm}(t)}}{(t+1)^{2}}\,dt.
  \end{align}
\end{subequations}
Therefore, Eqs.~\eqref{eq:Ey}, \eqref{eq:Vy}, \eqref{eq:y1}, and \eqref{eq:y2} are obtained.

\section{Zeros of the polynomial $g(y,h)$}

The zeros of the cubic polynomial $g(y,h)$ are
given by the Cardano formula:
\begin{equation}
  \label{eq:zeros.g}
  a_{1} = \omega\alpha+\omega^{2}\beta,\quad
  a_{2} = \omega^{2}\alpha+\omega\beta,\quad
  a_{3} = \alpha+\beta,
\end{equation}
where $\omega=e^{2\pi i/3}$ is a primitive cube root of unity, and
\begin{equation}
  \alpha^{3},\beta^{3} = \frac{h}{2} \pm \sqrt{\left(\frac{h}{2}\right)^{2} - \left(\frac{p\vphantom{h}}{3}\right)^{3}},\quad
  \alpha\beta = \frac{p}{3}.
\end{equation}
The discriminant of the polynomial $g(y,h)$ is
\begin{equation}
  \label{eq:discriminant}
  D = 108\left\{\left(\frac{p\vphantom{h}}{3}\right)^{3} - \left(\frac{h}{2}\right)^{2}\right\}.
\end{equation}
When $p\neq0$, a convenient parametrization is 
\begin{equation}
  \label{eq:h.mu}
  \frac{h}{2} = \left(\frac{p}{3}\right)^{3/2}\cos\mu,
  \quad\therefore
  \alpha,\beta = \left(\frac{p}{3}\right)^{1/2}e^{\pm i\mu/3}.
\end{equation}
Therefore, we have
\begin{subequations}
  \label{eq:zeros.mu}
  \begin{align}
    \label{eq:zeros.mu1}
    a_{1} &= -2\left(\frac{p}{3}\right)^{1/2}\cos\frac{\pi-\mu}{3},
    \\
    \label{eq:zeros.mu2}
    a_{2} &= -2\left(\frac{p}{3}\right)^{1/2}\cos\frac{\pi+\mu}{3},
    \\
    \label{eq:zeros.mu3}
    a_{3} &= \phantom{+}2\left(\frac{p}{3}\right)^{1/2}\cos\frac{\mu}{3}.
  \end{align}
\end{subequations}
We note that the parameter $\mu$ may be a complex number.

When the discriminant $D$ is positive, $g(y,h)$ has three distinct
real zeros.  Then, the order parameter $c(p,h)$ is given by
\begin{equation}
  c(p,h) = a_{3} - a_{1}
  = 2\sqrt{p}\cos\frac{\pi-2\mu}{6},\quad
  0\le\mu<\frac{\pi}{2},
\end{equation}
\textit{and} \eqref{eq:h.mu} when $h>0$.
If $h=0$, it is evident that $c(p,0)=2\sqrt{\max(p,0)}$.

\section{Solutions of the SDE \eqref{eq:SDE2} in the Symmetric Domain}

Herein, we introduce a positive parameter $k$ into the
function $g(y,h)$:
\begin{equation*}
  g(y,h) = -ky^{3} + kpy + kh.
\end{equation*}
If $h=0$, after some arithmetic, we can show that
\begin{subequations}
  \begin{align}
    y_{0}^{\pm}(t) &= \pm\sqrt{\frac{p}{1-\beta(t+1)^{-2kp}}},
    \quad
    \beta=1-4p,
    \\
    E(y_{1}^{\pm}(t)^{2}) &=
    \frac{1}{\{1-\beta(t+1)^{-2kp}\}^{3}}\left\{
      \frac{(t+1)^{-1}}{4kp-1} - \frac{3\beta(t+1)^{-1-2kp}}{2kp-1}
      - 3\beta^{2}(t+1)^{-1-4kp}
    \right.
    \notag
    \\
    &\quad\left.
      + \frac{\beta^{3}(t+1)^{-1-6kp}}{2kp+1}
      - \left(\frac{1}{4kp-1} - \frac{3\beta}{2kp-1}
        - 3\beta^{2} + \frac{\beta^{3}}{2kp+1}
      \right)(t+1)^{-4kp}
    \right\},
  \end{align}
\end{subequations}
for $p\neq0$, and
\begin{subequations}
  \begin{align}
    y_{0}^{\pm}(t) &= \frac{1}{2\sqrt{\tau(t)}},\quad
    \tau(t) = \frac{k}{2}\ln(t+1)+1,
    \\
    E(y_{1}^{\pm}(t)^{2})
    &= \left\{
      \left(1+\frac{3k}{2}+\frac{3k^{2}}{2}+\frac{3k^{3}}{4}\right)\tau(t)^{-3}
    \right.
    \notag
    \\
    &\quad\left.
      - \left(1 + \frac{3k}{2\tau(t)}
        + \frac{3k^{2}}{2\tau(t)^{2}}
        - \frac{3k^{3}}{4\tau(t)^{3}}
      \right)(t+1)^{-1}
    \right\},
  \end{align}
\end{subequations}
for $p=0$.
We note that the asymptotic behavior of the variance is estimated as follows:
\begin{equation}
  V(y^{\pm}(t))
  \sim 
  \begin{cases}
    \frac{(t+1)^{-1}}{4kp-1}, & kp>\frac{1}{4},
    \\
    (t+1)^{-1}\ln(t+1), & kp=\frac{1}{4},
    \\
    \left(\frac{1}{1-4kp} - \frac{3\beta}{1-2kp}
      + 3\beta^{2} - \frac{\beta^{3}}{2kp+1}
    \right)(t+1)^{-4kp},
    & 0<kp<\frac{1}{4},
    \\
    \left(1+\frac{3k}{2}+\frac{3k^{2}}{2}+\frac{3k^{3}}{4}\right)
    \{\ln(t+1)\}^{-3}, & kp=0,
    \\
    \left(
      \frac{\beta^{-3}}{4kp-1} - \frac{3\beta^{-2}}{2kp-1}
      - 3\beta^{-1} + \frac{1}{2kp+1}
    \right)
    (t+1)^{-2k|p|}, & -\frac{1}{2}<kp<0,
    \\
    (t+1)^{-1}\ln(t+1), & kp=-\frac{1}{2},
    \\
    \frac{(t+1)^{-1}}{-2kp-1}, & kp<-\frac{1}{2}.
  \end{cases}
\end{equation}

In \cite{Hod:2004,Huillet:2008,Hisakado:2012}, 
phase transition in a binary sequence system was discussed.
This system is a nonlinear P\'{o}lya urn with (in our notation)
the probability function
\begin{equation}
  f(z,0) = \frac{1}{2}\left(1 - \lambda\frac{1-2z}{1+t_{0}/t}\right),\quad
  -1<\lambda<1,
\end{equation}
where $t_{0}$ is initial time. When $t$ is large, we have 
\begin{gather}
  f(z,0) \simeq \frac{1}{2} + \lambda\left(z - \frac{1}{2}\right),
  \\
  \therefore
  f(z,0) - z \simeq (\lambda-1)y,
\end{gather}
where $y=z-1/2$ has been defined.
Thus, by keeping $kp=\lambda-1$ constant and taking the limit
as $k\rightarrow+0$, the variance converges to
\begin{equation}
  V(y^{\pm}(t))
  \sim 
  \begin{cases}
    \frac{1}{2\lambda-1}(t+1)^{2(\lambda-1)}, & -\frac{1}{2}<\lambda-1<0,
    \\
    (t+1)^{-1}\ln(t+1), & \lambda-1=-\frac{1}{2},
    \\
    \frac{1}{1-2\lambda}(t+1)^{-1}, & \lambda-1<-\frac{1}{2}.
  \end{cases}
\end{equation}
As $z$ is the proportion of $+$ (or $-$) symbols,
we introduce a scaled random variable $w=tz$ that represents
the number of $+$ (or $-$) symbols.
Its variance is
\begin{equation}
  V(w^{\pm}(t))
  \sim 
  \begin{cases}
    \frac{1}{2\lambda-1}(t+1)^{2\lambda}, & -\frac{1}{2}<\lambda-1<0,
    \\
    (t+1)\ln(t+1), & \lambda-1=-\frac{1}{2},
    \\
    \frac{1}{1-2\lambda}(t+1), & \lambda-1<-\frac{1}{2}.
  \end{cases}
\end{equation}
Thus, we conclude that
$\lambda=\frac{1}{2}$ is the critical point, and that $\frac{1}{2}<\lambda<1$
is a super-diffusion phase.

\end{document}